# Plasma-enhanced atomic layer deposition of titanium nitride for superconducting devices

John Femi-Oyetoro*, Sasha Sypkens, Henry LeDuc, Matthew Dickie, Andrew Beyer, Peter Day and Frank Greer

*Jet Propulsion Laboratory, California Institute of Technology, Pasadena, California 91109, USA*

This study presents a comprehensive investigation into the exceptional superconducting attributes of titanium nitride (TiN) achieved through plasma-enhanced atomic layer deposition (PEALD) on both planar and intricate three-dimensional (3D) structures. We introduced an additional substrate biasing cycle to densify the film and remove ligand residues, augmenting the properties while minimizing impurities. While reactive-sputtered TiN films exhibit high quality, our technique ensures superior uniformity by consistently maintaining a desired sheet resistance > 95% across a 6-inch wafer—a critical aspect for fabricating extensive arrays of superconducting devices and optimizing wafer yield. Moreover, our films demonstrate exceptional similarity to conventional reactive-sputtered films, consistently reaching a critical temperature ($T_c$) of 4.35 K with a thickness of around 40 nm. This marks a notable achievement compared to previously reported ALD-based superconducting TiN. Using the same process as for planar films, we obtained $T_c$ for aspect ratios (ARs) ranging from 2 to 40, observing a $T_c$ of approximately 2 K for ARs between 2 and 10.5. We elucidate the mechanisms contributing to the limitations and degradation of superconducting properties over these aggressive 3D structures. Our results seamlessly align with both current and next-generation superconducting technologies, meeting stringent criteria for thin-film constraints, large-scale deposition, conformality, 3D integration schemes, and yield optimization.

Titanium nitride (TiN) stands out as a material renowned for its dependable and versatile array of applications. Its exceptional properties empower the creation of robust, high-performance devices applicable in CMOS-compatible antennas[1], brain-inspired computing[2], quantum-dot optoelectronic devices[3] and superconducting detectors[4] addressing the stringent demands of today's technological landscape. In the realm of superconductivity, TiN showcases nonlinear kinetic inductance (KI) property[5], characteristic of a subset of transition metal nitrides. This attribute finds utility across diverse research disciplines, spanning high-coherence quantum processors[6], quantum transduction with extended lifetimes[7], ultra-sensitive photon detection[4,8], and quantum-limited parametric amplification[8,9]. Traditionally produced TiN films, used in the fabrication of superconducting devices (SDs) through reactive sputtering, demonstrate exceptional material purity[4,8]. This is primarily attributed to the utilization of a high-purity titanium target in the presence of reactive nitrogen gas.

*email: john.femi-oyetoro@jpl.nasa.gov   

This process bypasses unwanted precursor residuals and byproducts that are inherent in atomic layer deposition (ALD), reducing the likelihood of introducing a high level of impurities into the film. Nevertheless, ALD presents significant advantages, including atomic regulation of film thickness, high levels of uniformity and conformality, and improved resistance to pinhole and shadowing effects[10]. ALD-deposited TiN films with minimal impurities provide a dependable alternative to scarce sputtering targets. Their superior uniformity ensures homogenous material properties, critical for large-scale device fabrication and detector arrays, and increasing the yield of fully functional devices per wafer.

Obtaining high-quality superconducting TiN films via ALD presents challenges stemming from irregularities caused by contaminants, defects, random grain sizes and boundaries, and the inherent disorder associated with polycrystalline films. Precursor synthesis involves complex compounds, demanding strict measures to minimize impurities during deposition[10,11]. Complete impurity elimination is often unattainable[12–16]. Impurities can originate from other sources like vacuum system cross-contamination, low-purity process gases, and plasma source erosion, detrimentally impacting film conductivity[17,18] and superconducting properties[19,20]. TiN films utilized in SDs that exploit KI exhibit ≈ 100 μΩcm of resistivity, significantly higher than materials like aluminum (Al). This property minimizes microwave losses in the superconducting state, ensuring a high-quality factor in superconducting resonators and extended relaxation time, making TiN ideal for KI-based SDs[4,5,9,16,19–21]. While TiN films produced by ALD using halide precursors demonstrate high quality[19,20,22–24], metalorganic precursors like tetrakis(dimethylamido)titanium (TDMAT) offer advantages such as enhanced safety, increased reactivity leading to higher growth per cycle (GPC), and reduced process temperatures. The combination of plasma-enhanced ALD (PEALD) with TDMAT has shown limited but promising results, indicating the potential to bolster high-performance SDs[14–16]. While some of the aforementioned researchers have not extensively investigated the superconducting properties of ALD-deposited TiN, there remains sustained interest in SDs reliant on the KI of TiN.

In thin films, KI plays a significant role in high-frequency fields, as opposed to normal bulk metals or low frequencies where carrier collisions lead to rapid energy dissipation. At low temperatures, metal nitrides, such as TiN can enter a superconducting state, characterized by dissipationless KI nonlinearity driven by electron-phonon interactions[5]. This nonlinearity varies with current, described by the equation $L_k(0) \approx L_k(0)\left[1 + \left(I/I_*\right)^2\right]$, where $I_*$ characterizes nonlinearity (typically on the order of the critical current), and $L_k(0)$ is the linear inductance at low power. $L_k(0)$ can be approximated by sheet inductance $L_s$ with geometric factor $\gamma$, where $L_k(I) = \gamma L_s$. According to the Mattis-Bardeen theory[5,25], $L_s$ at low frequencies and $T = 0$ is given by $L_s = \hbar R_s/\pi\Delta$, with $R_s$ as normal-state sheet resistance and $\Delta$ denotes the superconducting band gap. This gap, tied to $T_C$, follows the relationship $2\Delta = 3.5 K_B T_C$. When incident photons exceed this energy threshold, they absorb into the material and momentarily break Cooper pairs, the superconducting pairs of electrons that find it more energetically favorable to enter a bosonic state at the Fermi level. TiN has a higher kinetic inductance fraction[4], $\propto \to 1$ as compared to Al[26], $\propto \approx 0.06$, allowing for extended quasiparticle recombination times with relatively lower material volume requirements. This necessitates the assessment of parameters such as $T_c$ and $L_s$ to assess the film's quality and its pivotal contribution in defining the fundamental superconducting properties of nonlinear KI devices.

Numerous SDs harness inherent KI found in transition metal nitrides as a foundational principle. These devices encompass, among others, microwave kinetic inductance detectors (MKIDs), kinetic inductance parametric amplifiers (KIPAs), and qubits. They share fundamental principles encompassing low dissipation, nonlinearity, superconducting transmission lines, resonators, and frequency selectivity. MKIDs are utilized for photon detection[4,5,8] and spectroscopy[27]. LeDuc et al.[4] pioneered the demonstration of highly sensitive arrays of microwave resonators (microresonators) utilizing TiN films. These MKIDs were fabricated through the process of reactive sputtering exhibiting remarkable sensitivity to dissipation signals. For far-IR, UV, and x-ray photon detection, the quasiparticle lifetimes extended reasonably long, up to 200 μs. The resonators also highlighted exceptionally low microwave losses, with internal quality factors ($Q_i > 10^7$), and the potential to achieve sensitivities below $10^{-19}$ WHz$^{-1/2}$. KIPAs[5,8,9] employ a distinctive traveling wave geometry to amplify signals, operating in close proximity to the quantum limit for parametric gain. Indeed intriguing, TiN has demonstrated low-loss microwave properties that are advantageous for high-coherence qubits, particularly with transmon qubits exhibiting coherence times of up to 60 μs on silicon[28] and 300 μs on sapphire[29] substrates. Moreover, employing superconducting TiN-coated through-silicon vias (TSVs) within quantum processors, characterized by high aspect ratios and substantial critical currents, presents the potential to alleviate interconnect congestion. This approach aims to maintain qubit coherence in a 3D architecture[6,30], demonstrating relaxation times of up to 12.5 μs.

Interestingly, film thickness inversely correlates with KI, highlighting the importance of precise film control. ALD proves ideal for these devices, ensuring uniformity, conformality, and large-scale deposition at the thin film limit. Limited yet substantial efforts have been directed towards the implementation of ALD techniques[6,16,19,20,31]. These initiatives have encountered significant challenges related to high process temperatures, achievement of high-quality films, corrosive precursor materials, as well as issues concerning, adhesion, repeatability and reliability. Elevated process temperatures can potentially lead to the huge formation of amorphous interfacial layers between the TiN film and the underlying substrate, which can introduce two-level systems (TLS)[32,33]. This, in turn, can result in high microwave losses and reduced $Q_i$, thereby constraining the performance of the devices. Loss mechanisms in KI-based SDs primarily vary based on design rules, predominantly occurring either within interfacial layers, material surfaces, or a combination of both, depending on the specific device design[21,34–36]. Concerns regarding reproducibility can arise due to tool variations, and issues with repeatability may manifest in multiuser tool environments. Therefore, addressing these challenges requires the adoption of a reliable ALD process operating at lower deposition temperatures while exhibiting satisfactory superconducting properties.

In this work, high-quality TiN films were deposited using PEALD with substrate biasing. Our primary goal was to maintain an exceptionally low concentration of impurities and achieve a film density close to nominal values. The objective is to attain outstanding superconducting parameters comparable to those obtained using well-established reactive sputtering techniques. Table I provides a summarized comparative overview (see supplementary material sections 1 and 2). This PEALD optimized process holds particular significance due to the intricate nature of ALD processes involved in conductive metal nitrides. Moreover, we successfully achieved conformal deposition within aggressive 3D structures, a distinct advantage inherent in ALD. This enhancement bolsters the feasibility of 3D integration and hybridization in SDs. Finally, we elucidated the mechanisms underlying the degradation of superconducting properties across these complex 3D structures. We employed two distinct sample thicknesses, approximately 40 nm and 80 nm, respectively. The selection of these thicknesses is based on prior device performance analyses and a specific emphasis on bulk properties.

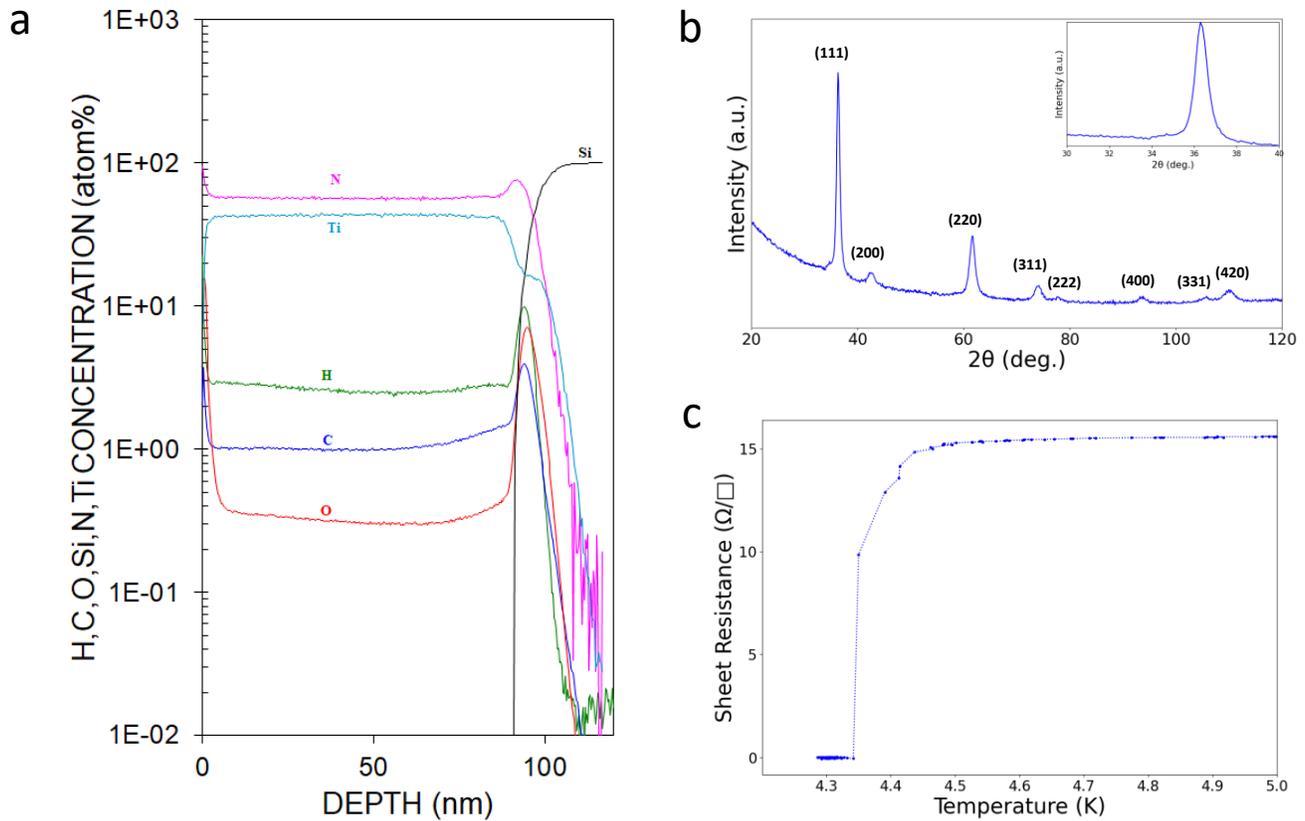

FIG. 1. (a) The SIMS spectra display contamination and composition profiles overlaid, revealing the atomic concentration levels of H, C, O, N, and Ti within an 80 nm PEALD TiN film. The TiN number density of 1E+23 atoms/cc serves as the basis for conversion into atom% units. (b) Grazing incidence XRD scan showing the phase identification. The inset depicts the preferential (111) orientation of the PEALD TiN film, (c) shows the resistance versus temperature measurement conducted on a 42 nm PEALD TiN film, demonstrating a $T_c$ of 4.35 K.

Film thicknesses within the 40-50 nm range have consistently yielded superconducting microresonators exhibiting high-quality factors in both reactive sputtered[4] and ALD-deposited TiN films[16,19,20]. Our films on 3D structures also align within this thickness range. Table 1 provides a comprehensive description of the properties of a 40 nm thick film, which has been deposited using both reactive sputtering and PEALD techniques. The crystal phase and structure of the sample appear to be consistent. However, it is noteworthy that the sputtered film exhibits a higher degree of crystallinity, evidently due to the utilization of a pure titanium target and pure $N_2$ gas during the deposition process. It's noteworthy that the superconducting properties exhibited by these films are comparable to those attained through reactive sputtering[4], and showcase superior metrics when contrasted with previously reported ALD-deposited TiN films[6,16,19,20]. The 80 nm-thick film complements this investigation by providing a comprehensive assessment of the material's bulk properties.

| Properties | PEALD TiN | Sputtered TiN |
|---|---|---|
| Stochiometric Ratio (Ti:N) | 0.95 | 0.98 |
| Mass Density (g/cm$^3$) | 5.12 | 5.12 |
| Crystal Orientation | <111> | <111> |
| Crystal Orientation (%) | 78.3 crystalline | 88.6 crystalline |
|  | 21.7 amorphous material(s) | 11.4 amorphous material(s) |
| Impurities concentration (%) | H: 2.5%, C: 1%, O: 0.3% | H: 0.66%, C: 0.025%, O: 0.1% |

Table I summarizes the mass density, crystal structure, and composition of both sputtered and PEALD TiN films.

In the scope of materials development, we utilized a substrate pre-treated with an HF-buffered solution and subsequently cleaned using Acetone, IPA, and DI water. The PEALD process was executed on a high-resistivity Si (100) substrate with resistivity > 10 kΩ cm. Employing TDMAT and a gas mixture of $N_2$/Ar, deposition occurred at a temperature of 300 °C. Additionally, we applied an average RF substrate bias voltage ranging between approximately 127-130 volts throughout the deposition process. This voltage aligns precisely with the optimal energy required to enhance film densification, eliminate ligand residues, and achieve the highest film quality. The studies in Ref. [15, 38] extensively examines ion energy as a function of bias voltage, closely matching our optimized bias voltage. To our knowledge, we are the first contributors to optimize this technique for fabricating high-quality superconducting TiN. Each ALD cycle resulted in a deposition rate of approximately 0.65 Å, enabling meticulous and uniform control of film thickness across the entire 6-inch wafer. This is more evidenced by the sheet resistance map, showcasing a level of uniformity > 95% via eddy current contactless measurement. The surface roughness of 1 µm × 1 µm regions on a 40 nm thick film was 1 nm (See supplementary material FIG S1 and FIG. S3).

To achieve a comprehensive understanding of the quality of our PEALD TiN films, we conducted an extensive material characterization employing various techniques, including atomic force microscopy (AFM), scanning electron microscopy (SEM), contactless sheet resistance measurements, cryogenic DC electrical measurements, X-ray photoelectron spectroscopy (XPS), X-ray diffraction (XRD), X-ray reflectometry (XRR), secondary ion mass spectrometry (SIMS), transmission electron microscopy (TEM), energy-dispersive X-ray spectroscopy (EDS), and electron energy loss spectroscopy (EELS). Film thicknesses were determined through XRR and TEM. XRR is also utilized to assess mass density and surface roughness, while AFM was employed to validate the film's roughness. For the superconductivity measurements, we employed a specialized cryogenic system cooled by liquid helium (LHe). Additionally, we utilized a current source and voltmeter apparatus for measurements.

Figure 1 (a) shows a SIMS spectrum to quantify impurity levels, given its heightened sensitivity to low elemental quantities. The H, C, and O content were found to be approximately 2.5%, 1%, and 0.3%, respectively. The XPS depth profile conducted within the bulk material revealed an average Ti:N ratio of 0.95. XRD was employed to identify the sample's phase and crystal structure through fiber texture. The phase information obtained serves as a semi-quantitative measure of crystallinity, as outlined in Table 1 and shown in Figure 1 (b), the lattice constant measures 4.283 Å, with a bond length of 2.142 Å. The film demonstrates a heightened level of texture alignment towards the <111> crystallographic direction, matching the orientation reported in Ref. [15]. (see supplementary data). To investigate the superconducting properties of the planar samples, a 6-inch wafer of PEALD TiN was precisely diced into individual chips. Each chip underwent patterning with van der Pauw structures and subsequent attachment and wire-bonding to a carrier chip. These chips were integrated into a cryogenic system cooled by LHe and utilized a specialized dipstick apparatus. Figure 1 (c) illustrates a 42 nm film, measured using XRR and TEM, displaying a $T_c$ of 4.35 K. Notably, our film's superconducting characteristics are comparable to those achieved through reactive sputtering[4] and exhibit superior metrics compared to previously reported ALD-deposited TiN films[6,16,19,20,31,37].

To comprehensively characterize the planar and intricate 3D structures, the combination of TEM, EDS, and EELS were employed for crystal structure and size, and compositional analysis of the films. This technique, in conjunction with superconductivity measurements, aids in elucidating the film's performance within 3D structures characterized by varying aspect ratios. Figure 2 (a) shows a BF HRTEM of an 80 nm planar film, with multiple columnar grain structures observed, and grain widths reaching approximately 60 nm. Very smooth final film outcome with what seems to be a multilayer stack comprising of $SiN_x$, TiN nucleation layer, and TiN "bulk" film. In addition to Figure 2 (a), Figures 2 (c) and (e)-(g) reveals the existence of an approximately 2 nm-thick silicon-nitride interface layer positioned between the TiN film and the Si wafer using EELS and EDS.

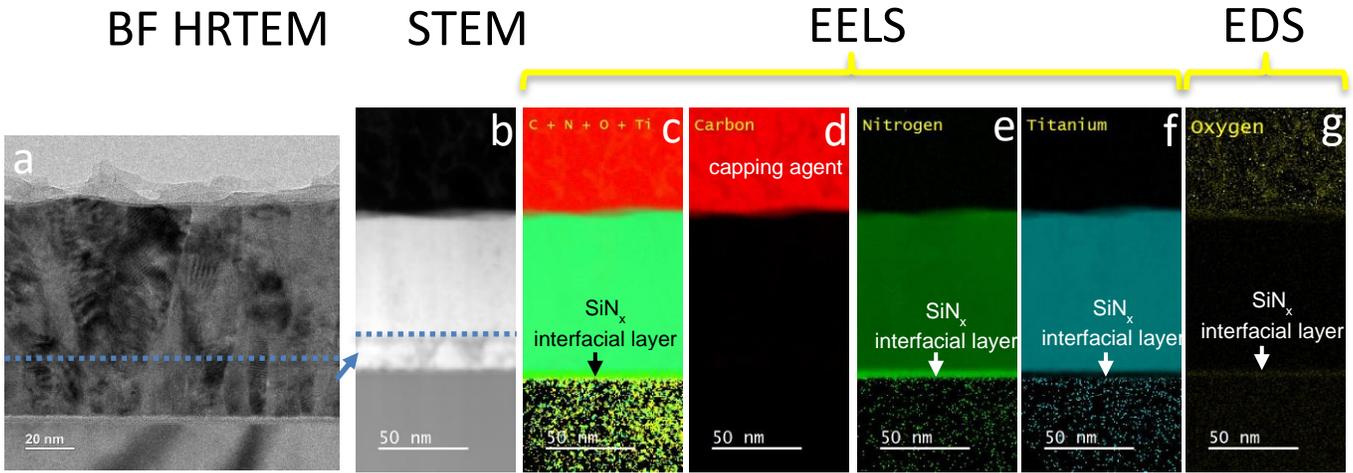

FIG. 2. (a) depicts a bright-field HRTEM image. Below the delineated blue dash lines, smaller grains are observed, predating the emergence of larger grains visible above the dash line, as depicted in (b). The high-angle annular dark-field STEM image in (b) highlights contrast variations within the multilayered grain boundaries, (c-f) presents EELS data for C, N, Ti, and N revealing a $SiN_x$ interfacial layer, (g) presents EDS data detecting very minimal oxygen content within the PEALD TiN films.

The observed incubation period is likely the result of the initial nucleation triggered by exposure to nitrogen plasma. This initiation subsequently supports the nitridation process of the silicon. Such a phenomenon is prevalent in nitrogen-based TiN growth chemistries[8,21] unlike gas chemistries that do not involve nitrogen[15,38].

Nine DRIE trenches were utilized instead of vias to streamline characterization procedures for 3D structures. Trenches offer the possibility of conducting measurements from the front side of the sample. The utilization of 3D architecture in SDs is highly advantageous in addressing interconnection crowding and enabling hybridization. In Ref [37], TiN was used as superconducting TSVs to effectively integrate transition edge sensors with SQUID readouts. Additionally, in Ref [6, 30], a similar approach was utilized to route signals to resonators, compacting the device while maintaining coherence. The trench aspect ratio (AR) we utilized varied between approximately 2 and 40 to establish future definitive design guidelines for superconducting TSVs. Thermal oxidation and HF wet etching were employed to smoothen the sidewalls, eliminating the DRIE scalloping (see supplementary material section 3). Subsequently, TiN was deposited within the trench structures using the same process as the 40 nm-thick planar film. Figure 3(a) displays the mask utilized in the DRIE process to fabricate the 9 trenches. Figures 3(b) and (c) present high-resolution SEM images of Trenches 1 and 9, respectively with a conformal PEALD TiN film. Figure 3(d) exhibits an SEM image showcasing all trenches deposited with the PEALD TiN film post a comprehensive DRIE process. Figures 3 (e)-(k) features Trench 9 which has the highest AR of 40. In Figure 3 (e) and (f) the BF

HRTEM image presents a film with a smooth texture and a rough morphology at the corner akin to that of Trench 1, albeit less rough than that of Trench 7 (see supplementary material section 4). A slight oxide layer is discernible, primarily concentrated at the corners, aligning with similar characteristics found in the prominent corners of Trenches 1 and 7 (see supplementary materials section 4). The films along the sidewalls in Figure 3 (h) demonstrate a slightly rougher texture and randomly oriented grains compared to the top surface in Figure 3 (f). There is an absence of an interface layer, and a minute presence of $TiO_2$ is noted, featuring a composition notably rich in nitrogen. A significantly thick oxide layer is evident at the base in Figure 3(i) indicates the consequence of the high AR and the inability of the post HF-treatment to reach the bottom within the necessary timeframe. This contrasted with Trenches 1 and 7, where this particular issue did not manifest (see supplementary materials). Figures 3(j) and (k) respectively present EDS maps of (f) and (i), revealing a well-balanced Ti/N mixture.

We utilized a laser trimming technique to remove TiN from the edges and lip of the trenches, followed by laser cleaving of the opposite edge of the sample, as illustrated in the Figure 4 (a). This technique ensures the forced direction of current through the trenches, bypassing the planar TiN, which exhibits the least resistance. If this method had not been employed, we measured a $T_c$ of approximately 4.35 K, similar to that of the planar 40 nm sample. Superconductivity measurements were conducted separately on each trench and between trenches 1 and 9, totaling 10 individual measurements for assessment of TiN conformal superconductivity. Two contacts were positioned at either side of each trench for a four-wire measurement setup. The current path followed from the field, traversed down one sidewall, along the bottom, and returned up the opposite sidewall to the other side. During measurement, the temperature exhibited a steady decline from approximately 3K to 100mK within a 30-minute duration. A compact radiation shield was utilized to shield the trenches, mitigating the impact of 3K radiation on the film. Overall, this approach aimed to characterize the $T_c$ for the TiN within each trench, identifying the points of superconducting degradation and elucidating the underlying mechanisms. The Ic across the trenches was not measured due to the power requirement for the $I_c$ measurement exceeding the cooling power of the mK stage. Nevertheless, we are actively exploring robust measurement methodologies tailored specifically for high $T_c$ TSVs in integrated SDs.

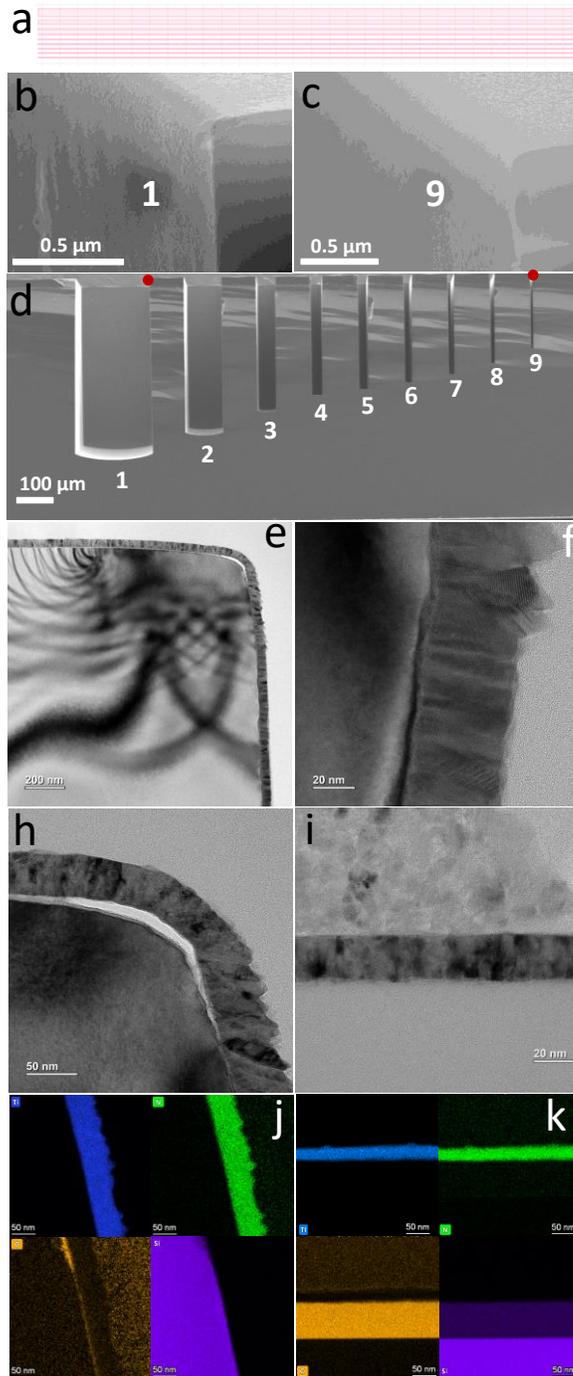

FIG. 3 (a) Illustrates the mask employed for trench patterning, not drawn to scale. The mask features trench widths of 200 µm, 100 µm, 50 µm, 30 µm, 25 µm, 20 µm, 15 µm, 10 µm, and 5 µm, evenly spaced at 100 µm intervals. High-resolution SEM images in (b) and (c) correspond to the delineated red areas in (d), representing trenches 1 and 9, respectively. The labeled trenches 1-9 in (d) exhibit aspect ratios achieved through deep reactive ion etching, measuring 2, 4, 6.5, 9.5, 10.5, 15, 20, 30, and 40, respectively. (e) shows a BF HRTEM image of the flat top corner and sidewall of trench 9, representing a highly conformal film. (f) exhibits a high magnification BF HRTEM image of the top corner of trench 9 where the underlying residual oxide from the trench smoothing cycle is visible, with a reduced TiN film. (g) displays a high magnification BF HRTEM image of the sidewall of trench 9, exhibiting elongated material and misaligned grains. (i) presents a high magnification BF HRTEM image of the bottom of trench 9, showing a more pronounced underlying oxide and significantly less TiN film. (j) and (k) respectively depict EDS maps of Ti, N, O and Si of (f) and (i) respectively.

As shown in Figure 4 (b), all nine trenches exhibited superconducting behavior with polycrystalline grains at corners, sidewalls and bottoms, underscoring the robustness of the PEALD TiN process. We also probed in-between Trench 1 and 9, which continues to display superconductivity. In general, TEM highlighted peculiarities in step coverage (see supplementary section 4). While the coverage appeared highly conformal along the sidewall, differences in film morphology were noted at the top corners of Trenches 1, 7 and 9, likely attributed to residual oxide from trench smoothing cycles. Additionally, grain rotation was observed at various locations within the top corners and top sidewalls. This distinct variations in the film morphology are evidently based on the incident angle of the ion flux.

We assess the step coverage and composition within Trenches 1, 7 and 9 using TEM, EDS and EELS (see supplementary material sections 2.5 and 4), and its impact on $T_c$ of the coated TiN. At differing angles, the TiN film exhibited faceting and grain orientation changes, influencing the film thickness and roughness as shown in Figures 4 (c)-(g). Notably, the corners and sidewalls in Figure 4 (d) and Figures 4 (e)-(f) respectively exhibit reentrant material, elongated, and misoriented grains, contributing to an overall rougher surface characteristic. Trenches 1-6 exhibit typical transitions around 2 K, whereas Trench 7 displayed a $T_c$ breakpoint, indicated in Figure 4 (b), suggesting the occurrence of multiple transition levels at lower temperatures. This observation distinctly denotes the consequential alterations in film properties, including morphology and composition. Alongside the resistance imposed by the trench geometry[30], disruption of the columnar growth morphology contributes to the degradation of the superconducting state. Understanding and characterizing the impact of degradation in superconductors are crucial for optimizing the performance of SDs. This comprehension provides critical insights into the fundamental mechanisms governing superconductivity and aids in devising strategies to mitigate adverse effects, such as pair-breaking, quasiparticle scattering, inhomogeneous current and energy dissipation. A disordered lattice structure can disrupt the coherent motion of superconducting electrons, affecting the formation and stability of Cooper pairs, consequently reducing $T_c$ and weakening the superconducting state.

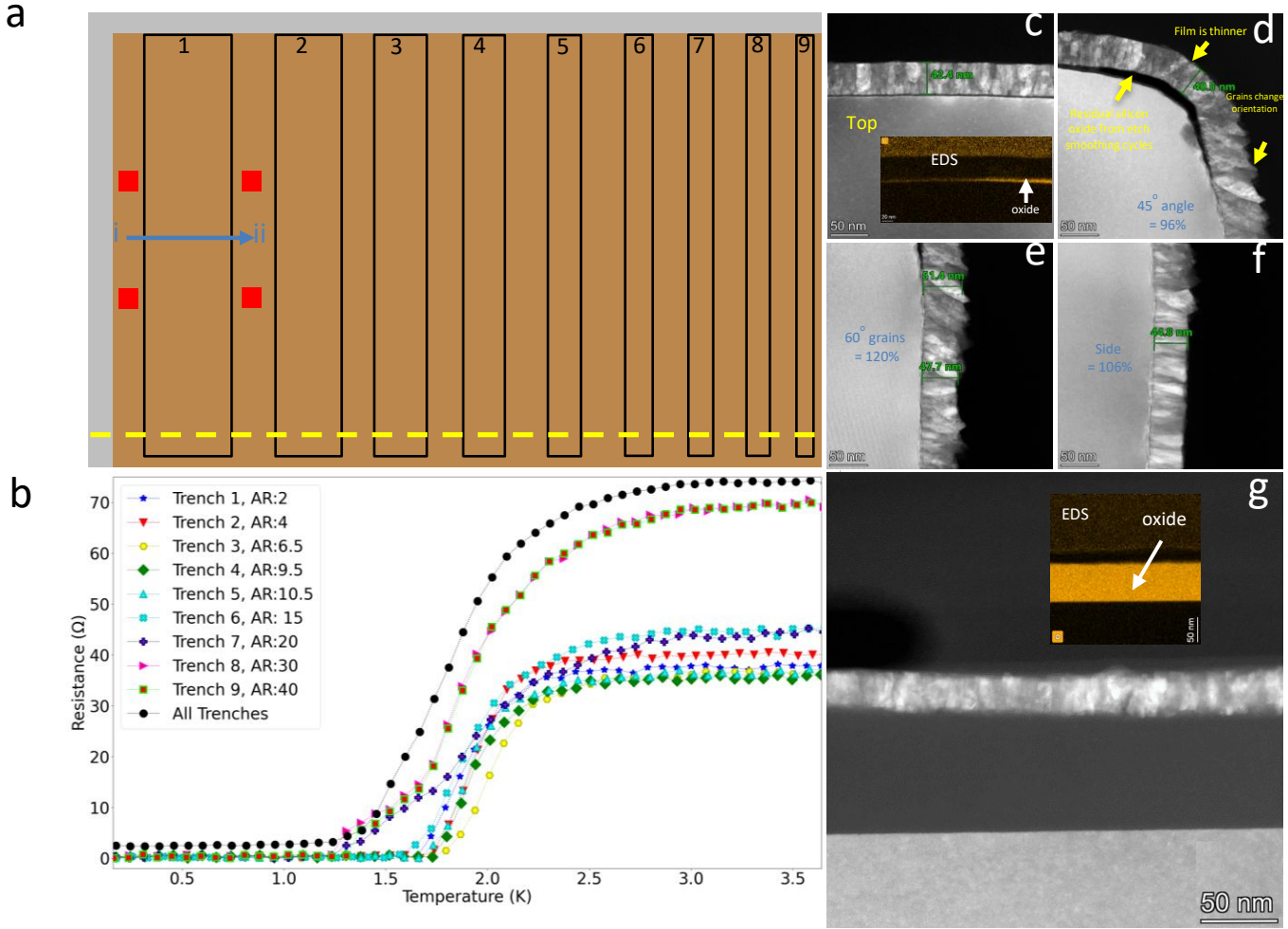

FIG. 4. (a) presents a schematic outlining the laser trimming procedure used for eliminating TiN from trench edges and lips, not drawn to scale. The delineated black lines denote trenches numbered 1-9 from left to right. The golden-colored region represents TiN, present both on the planar surface and within the trench, while the grey area indicates the region where TiN was removed using laser ablation, revealing the Si substrate. The dashed yellow line on the opposite edge of the sample signifies the laser-cleaved region. Red pads exemplify the positions where four-wire connections were bonded for measurements. The blue arrow, running from i to ii, illustrates the enforced current direction flowing within the trench due to the electrical isolation imposed by laser trimming and cleavage, (b) the plot depicts resistance versus temperature measurements, emphasizing the critical temperature $T_c$ for Trenches 1-9, as well as for the region between Trenches 1 and 9 (All Trenches), with scaling from 700 Ω, (c)-(g) shows variations in morphology, grain orientations and composition in planar PEALD TiN film of Trench 9 in (c), top corner in (d), sidewalls in (e) and (f) and at the bottom in (g)

This study and prior research highlighted the impact of ion energy on film growth[39–41]. We attribute the "breaking" mechanism to the behavior of ions ($N_2^+/Ar^+$), predominantly normal to the substrate, resulting in varied ion and radical transport to the trench bottom, particularly pronounced in higher AR trenches. Ions flux to sidewall is much less than field and at a glancing angle. The TDMAT precursor has cosine distribution and the flux survives bounces. For instance, the reactive radicals ($N^*$) have cosine distribution and radical flux decreases such that: $N^*$ + N(surface) → $N_2$(g) leading to recombination. Future experiments may involve KOH etching of silicon to determine the impact of sidewall angle on film characteristics (see supplementary section 5). Ion densification of the film and faceting the corners of the trenches could also potentially enhance the superconducting properties. The orientation of grains relative to the feature bottom and surface normal appears to be influenced by both ion behavior and AR, suggesting their significant roles in film morphology and properties. The ion bombardment via biasing could also lead to grain anisotropic from planar down to the bottom of the trench. On planar surface, the grains are vertically aligned from the directional ion bombardment. However, at the corners the change in directions of the ions leads to misalignments of these grains. This phenomenon is not observed at the lower sidewalls and bottoms of the trenches.

In summary, we present an optimized PEALD technique designed for depositing high-quality superconducting TiN. We observed both conformal deposition and the retention of superconducting properties on highly aggressive 3D trenches, up to AR of 40. PEALD faces limitations attributed to the finite lifetime of radicals and charged species. Although thermal ALD is known for its ability to coat highly conformal structures with high ARs up to two orders of magnitude better than PEALD[11] due to enhanced directionality, a trade-off arises in the context of superconducting materials that require high $T_c$. Prior reports have employed substantial thicknesses of TiN (150-230 nm), elevated process temperatures, or circumvented the use of a PEALD method[6,30,37]. Our findings demonstrate that, with a reduced thickness of 4 orders of magnitude, we have achieved comparable $T_c$. Indeed, since $T_c$ scales directly with film thickness, we propose that thicker films in vias will yield even higher critical currents and $T_c$ values. Evidently, there exists a trade-off between maintaining consistent morphology and composition in high ARs and ensuring material purity. In this scenario, material purity seems to take precedence. It is clear that if grains are oriented towards feature bottom – ions are important, and if grains are normal to the surface – AR is important, and the $T_c$ appears to start breaking at an AR of 20.

## ACKNOWLEDGMENTS

This research was primarily conducted at the Jet Propulsion Laboratory (JPL), California Institute of Technology, under contract with the National Aeronautics and Space Administration (NASA). It also involved work at the Kavli Nanoscience Institute, and the Molecular Materials Research Center in the Beckman Institute of the California Institute of Technology, as well as contributions from Eurofins EAG Laboratories for XRD and SIMS analyses, Covalent Metrology for TEM, EDS, and EELS investigations, and Laserod Technologies, LLC for laser ablation procedures.

## COMPETING INTERESTS

The California Institute of Technology has filed a U.S. utility patent titled 'Atomic Layer Deposition of Superconducting Transition Metal Nitrides for Quantum Circuits and Detectors' (inventors: J.F.O and F.G.), which includes the methods and technologies related to the research discussed in this publication.

**Supplementary Information:**

**Plasma-enhanced atomic layer deposition of titanium nitride for superconducting devices**

John Femi-Oyetoro, Sasha Sypkens, Henry LeDuc, Matthew Dickie, Andrew Beyer, Peter Day and Frank Greer

*Jet Propulsion Laboratory, California Institute of Technology, Pasadena, California 91109, USA*

S1. **Materials Development of PEALD TiN**

TiN films were successfully deposited on a 6-inch high-resistivity intrinsic/undoped Si (100) substrate with a resistivity > 10 kΩ ohm-cm. The substrate used was a 675±15 µm prime float zone (FZ) wafer obtained from WaferPro LLC. It exhibited a single semi-flat bow/warp ≤ 30 µm, and a total thickness variation (TTV) ≤ 5 µm. The wafer was also required to have ≤10 particles that are ≥ 0.3 µm. Prior to deposition, the wafers underwent a cleaning process consisting of sequential treatments with acetone, methanol, and propanol, followed by thorough rinsing with de-ionized water. Oxford Flex II PEALD system was utilized using TDMAT as the precursor and a nitrogen/argon mixture for the plasma step. The PEALD reactor consists of a water-cooled copper coil wrapped around a cylindrical alumina tube and is connected to a radio frequency (RF) power supply operating at 13.56 MHz with a maximum power of 600 W. This inductively coupled plasma (ICP) source generates radicals and ions during the plasma exposure step. In the FlexAL configuration, an external RF power supply operating at 13.56 MHz and up to 100 W is connected to the reactor table, enabling substrate biasing with fully automated RF matching. The precursor temperature was set to 60 °C, and the pulse period for the deposition cycle was at 300 ms. During the TDMAT half-cycle, the chamber pressure was 220 mTorr and 3mTorr for the plasma half-cycle. The argon flow rate through the precursor cannister was maintained at 30 sccm, and 10 sccm during the plasma half-cycle. The precursor delivery line was heated to 100 °C to prevent condensation. High-purity nitrogen and argon gases (>99.999% purity) were used to generate the plasma, with a flow rate of 10 sccm each. The gases were stabilized for 5 s before plasma ignition with pulse periods of 20 s, a power of 300 W, an intermittent average bias voltage step of -127 V, at a deposition temperature of 300°C. At this temperature, the substrate table was maintained for preheating while flowing argon gas at a rate of 200 sccm for 10 mins, and precondition step with 25 cycles before actual deposition.

To enhance film quality, an ion bombardment intermittent step was introduced by implementing substrate biasing during the final half-seconds of plasma exposure. The bias was applied in the last 10 seconds of the 20-second total plasma exposure. Typically, the latter half (10s) of the plasma exposure time is considered optimal to prevent the simultaneous ignition of the plasma and bias, which can result in the breakage of ligand species. This methodology serves to enhance film densification, eliminate precursor residues, and facilitate the diffusion of surface adatoms. Previous investigations have demonstrated that maintaining low pressure effectively reduces sputtering in ICP tubes, with a more pronounced impact observed in quartz tubes as opposed to alumina tubes. Throughout the plasma exposure, there is a conspicuous reduction in the concentration of dissociated nitrogen, resulting in comprehensive nitridation. Ion bombardment resulting from substrate biasing exerts a substantial influence on the growth, microstructure, and composition of the TiN films. This process induces modifications in the film's properties, attributed to enhanced adatom mobility due to momentum transfer imparted by the bombarding ions, highlighting multi-grain sizes.

S2. **Materials Characterization**

S2.1 **X-ray Diffraction and X-ray Reflectivity**[1]

The samples were mounted directly to the diffractometer with a vacuum chuck sample stage. Grazing incidence XRD (GIXRD) data were collected using a Bruker D8 UltraGID 7-axis diffractometer equipped with a copper X-ray tube, parallel beam optics, a 0.5° parallel slit analyzer and a scintillation detector. Grazing angles of 0.36 and 0.38 degrees were used for Sample1-40nm and Sample2-80nm, respectively. For X-ray reflectivity (XRR), the samples were mounted directly to the diffractometer with a vacuum chuck sample stage. data were acquired with a Bruker D8 Discover UltraGID 7-axis diffractometer equipped with a copper X-ray tube, parallel-beam optics, a narrow XRR receiving slit and a scintillation detector.

S2.2 **Phase Identification, Semi-Quantitative Analysis and Crystallite Orientation (Simple Texture)**[1]

Crystalline phases are distinguished by analyzing the positioning and relative intensity of peaks within experimental X-ray diffraction (XRD) data, comparing them to entries in the ICDD/ICSD diffraction database. Reference markers within the phase serve as indicators of where the anticipated experimental peaks should be situated on the two-theta scale, while the markers' relative heights denote the expected peak intensities for a fine-grained and randomly oriented phase. It's crucial to note that while XRD is sensitive to crystal structure, it has limited sensitivity to elemental or chemical state compositions. Hence, the actual chemistry of the matching phase may differ from that of the reference card.

Identifying amorphous phases differs from crystalline phases due to their lack of distinct, sharp peaks. Instead, amorphous phases exhibit a few broad peaks, approximately 100 times wider than a crystalline peak. Additionally, the positioning of an amorphous peak is determined based on short-range nearest neighbor distances, primarily among elements with the most electrons, rather than the material's composition. Consequently, XRD cannot ascertain the composition of amorphous materials. Semi-quantitative analysis involves employing WPF (whole pattern fitting), a subset of Rietveld Refinement that incorporates all intensity above the background curve. This technique necessitates knowledge of either the structure factors and atomic locations or the reference intensity ratio for all identified phases. During this procedure, parameters such as structure factor (linked to concentration), lattice parameters (associated with peak position), peak width, and peak shape are refined for each phase to minimize the R value—an estimation of the agreement between the model and experimental data across the entire pattern. The WPF representations comprise a difference curve displayed at the pattern's top, demonstrating the discrepancy between the experimental data and the modeled pattern. This difference curve is represented in square-root intensity to highlight minor dissimilarities between the patterns.

If an amorphous phase registers within the XRD pattern, it necessitates assigning a density to quantify its presence accurately. Any discrepancy in this density directly impacts the quantification of the amorphous material. Consequently, while the relative concentrations of crystalline phases remain accurate, the absolute concentrations incur an error proportional to the inaccuracy in the amorphous concentration.

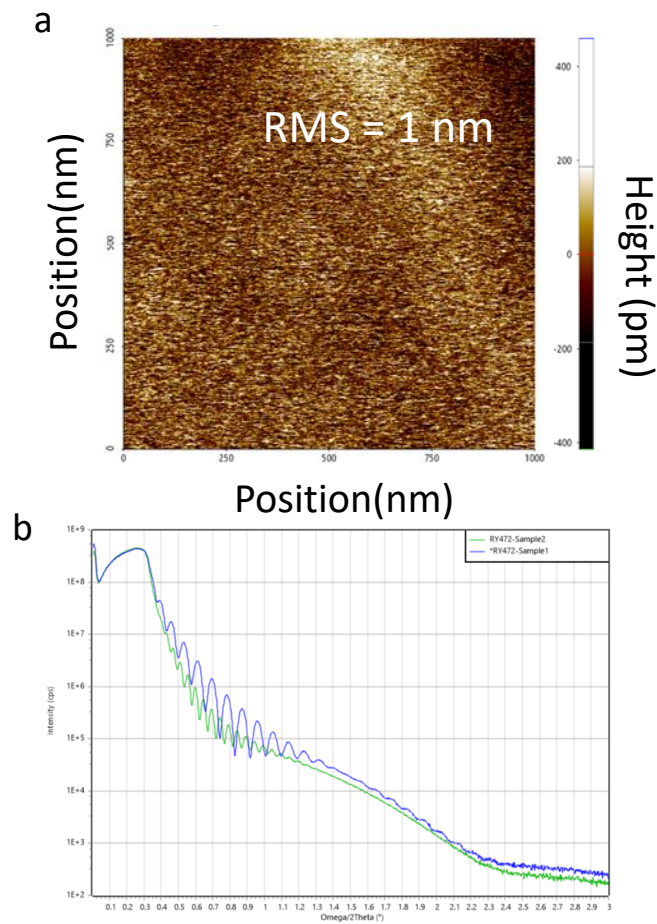

FIG. S1. (a) AFM of a 40 nm PEALD TiN film, with an RMS of 1 nm (b) shows the XRR data comparing a 40 nm and 80 nm film.

It's important to note that the size of the amorphous peak significantly relies on the background's shape in this region, and the assignment of the background is subjectively determined by the analyst. The determination of crystallite orientation involves collecting XRD data in a symmetrical theta/two-theta geometry to examine crystallographic planes parallel to the sample's surface. The peak ratio method is employed to ascertain crystallite orientation, utilizing the areas of multiple peaks from diverse crystallographic directions. These intensities are adjusted concerning that of a randomly oriented phase, obtained from the ICDD/ICSD database. Each peak undergoes profile fitting, and the resulting area is utilized to derive intensity ratio values recorded in a summary table. These values represent the percentage of crystallites oriented in three or more distinct crystallographic directions. The peak with the least contribution to the intensity is presumed to originate from randomly oriented grains. This contribution is subtracted from the totals of other peaks to establish the random fraction. Profile fitting is employed to ascertain peak positions, widths, and areas, crucial for calculating crystallite orientation.

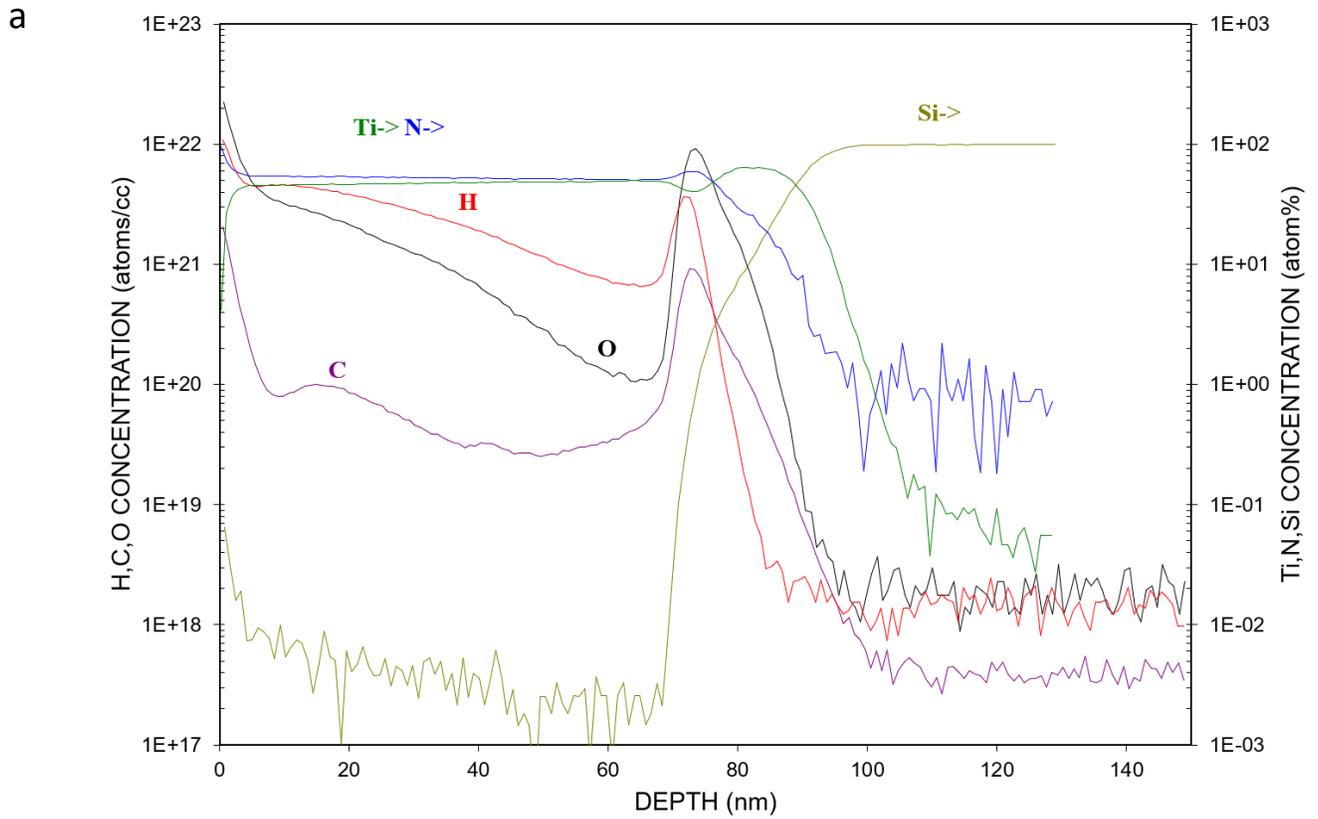
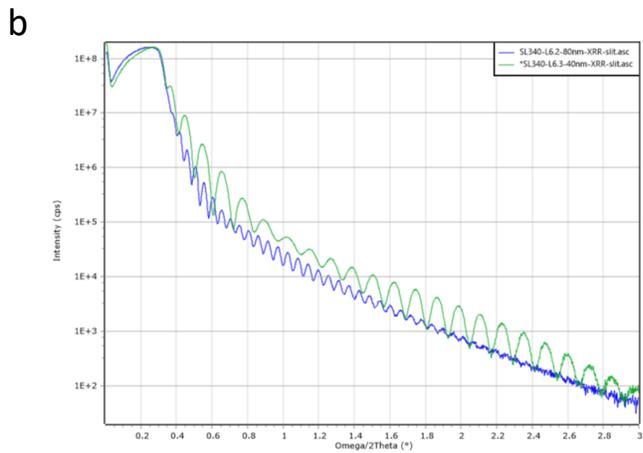
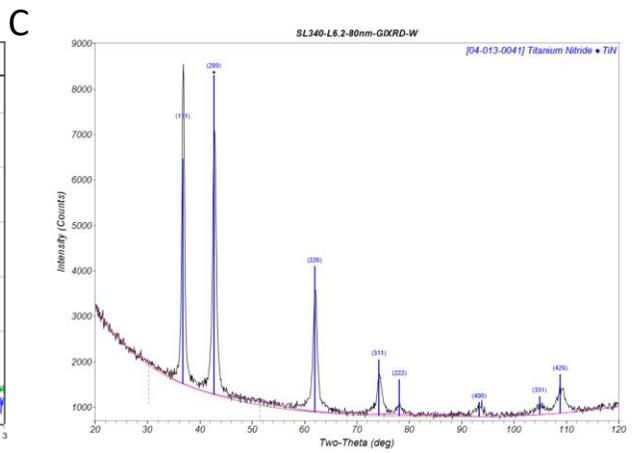

FIG. S2. (a) SIMS spectra display contamination and composition profiles revealing the atomic concentration levels of H, C, O, N, and Ti within an 80 nm reactive sputtered TiN film. (b). shows the XRR data comparing a 40 nm and 80 nm film, (c) shows grazing incidence XRD scan showing the phase identification and a preferential (111) orientation of film.

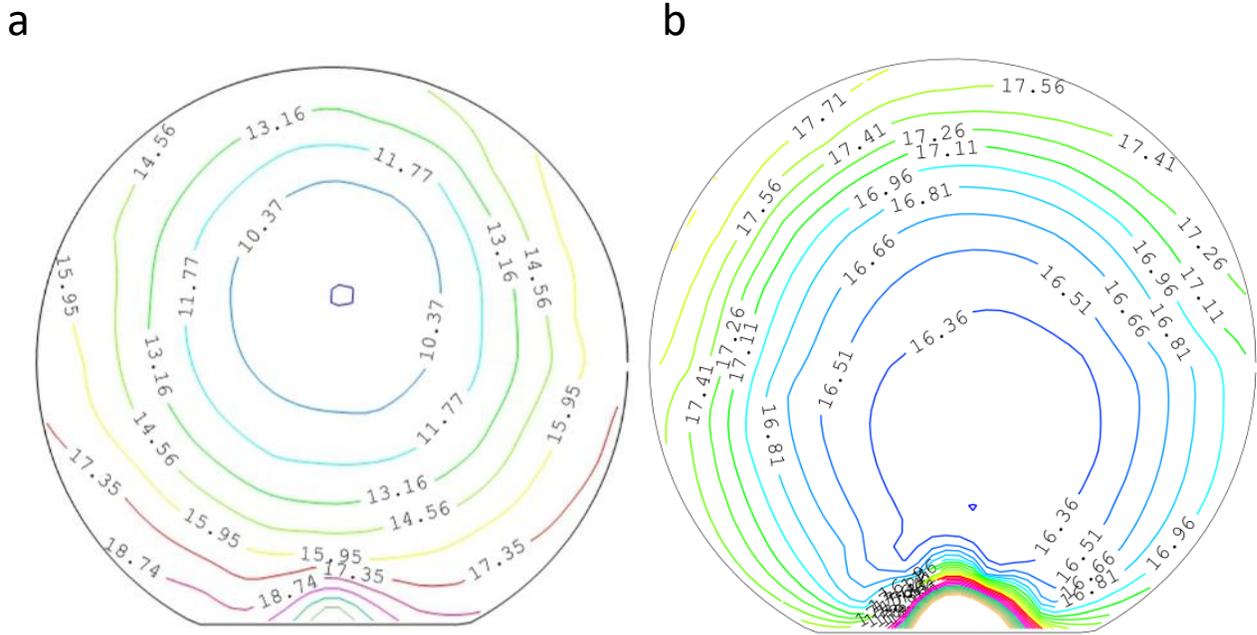

FIG. S3. displays the eddy current contactless sheet resistance of (a) reactively sputtered TiN and (b) PEALD TiN. The 103-point map was measured, excluding a 0.1-inch edge, across a 6-inch wafer. The uniformity levels are 77% for (a) and 95% for (b).

S2.3 **X-ray Photoelectron Spectroscopy**

XPS measurements, a Kratos Axis Ultra system with a monochromatic Al Kα source (hν = 1486.6 eV) operating at 150 W and a base pressure of < 1.0 x $10^{-9}$ Torr was utilized in the analysis chamber. The X-rays were angled at 45°, and the hemispherical analyzer collected photoelectrons at 90° to the sample surface. The spot size on the sample was approximately 700 µm x 300 µm. High-resolution spectra were acquired at an operating current of 10 mA, voltage of 15 kV, and a resolution of 50 meV with a pass energy of 10 eV for each sample. The spectra were aligned by referencing the C 1s (284.6 eV) transition and fitting the Ti 2p, N 1s, O 1s, and C 1s peaks. To perform depth profiling, a 2 kV $Ar^+$ milling process was carried out with a raster size of 3 mm x 3 mm for 10 minutes, resulting in the removal of less than 45 nm from the surface of the TiN samples. The instrument was operated using Vision Manager software v. 2.2.10 revision 5, and the spectra were analyzed using CasaXPS software (CASA Software Ltd).

S2.4 **Secondary Ion Mass Spectroscopy**[1]

The SIMS analysis employed a Phi Adept 1010 quadrupole tool with a primary ion energy of 2 keV. Contamination profiles, specifically for hydrogen, carbon, and oxygen, were acquired using negative atomic ions. Composition profiles for titanium, nitrogen, and silicon were obtained using positive cesium cluster ions.

S2.5 **TEM, EDS and EELS**[2]

The specimens were coated with protective carbon, sputtered with Pt, as well as e-W and i-W. Lamellae were fabricated employing the lift-out preparation methodology, using the Thermo-Fisher (FEI) Helios UC FIB-SEM system and the Thermo-Fisher (FEI) Helios UX FIB-SEM. Transmission electron microscopy (TEM) images was captured using a JEOL JEM-F200 Multi-Purpose Electron Microscope, which was operated at an acceleration voltage of 200 kV and was equipped with a Gatan OneView CCD camera. Energy-dispersive X-ray spectroscopy (EDS) data was collected via a Dual JEOL JED-2300 Dry SDD EDS detector, and electron energy loss spectroscopy (EELS) data was gathered with the GIF Continuum ER detector. Subsequently, the acquired data underwent processing using the DigitalMicrograph® and Velox® software packages.

S3. **Process and Cryogenic DC Characterization of Planar and 3D structures**

To access the superconducting properties, the TiN samples deposited with SPR 220-3.0 photoresist and spin-coated at 3000 rpm for 30 s, followed by a soft bake at 115 °C for 90 s. Lithographic patterning was executed without a mask, in which van der Pauw structures were defined, employing the MLA-150 system. The applied exposure dose was 350 mJ/cm$^2$ at a laser wavelength of 405 nm. AZ 300 MIF was used for development. Afterwards, a reactive-ion etching (RIE) process was implemented using a mixture of $BCl_3/Cl_2/Ar$ gases at flow rates of 30/30/10 sccm, and at a pressure of 5 mTorr. The plasma was generated with an ICP RF power of 400 W and a bias RF power of 50 W. This was followed by a post-etching treatment involving an $O_2$ plasma step. To complete the process, the resist material was removed using a combination of acetone, isopropyl alcohol, and deionized water. The prepared samples were then diced into individual chips using a Disco DAD 320 dicing saw, and wire-bonded onto a carrier chip. The superconductivity parameters were assessed using a current source (Keithley 224) and a voltmeter in a cryogenic system cooled by liquid helium (LHe), utilizing a specialized dipstick apparatus.

For the trenches, the wafers underwent two RCA cleaning procedures. The first RCA clean step was carried out using 4200 ml of DI water, 2520 ml of $NH_4OH$, and 840 ml of $H_2O_2$ at a ratio of 5:3:1. Subsequently, the second RCA clean step was conducted, involving 5040 ml of DI water, 1260 ml of HCl, and 1260 ml of $H_2O_2$ at a ratio of 4:1:1, all performed at 80 °C for 10 minutes. Following the cleaning steps, an AZ 10XT-520cP photoresist was applied through spin-coating at 2000 rpm for 40 s, followed by a soft bake at 115 °C for 120 s. The wafers were then subjected to a rehydration hold lasting for 1 hr. Lithographic patterning was executed without the use of a mask, utilizing the MLA-150 system. This process involved an exposure dose of 500 mJ/cm$^2$ at a laser wavelength of 405 nm. Subsequently, the pattern was developed using a 1:4 mixture of AZ 400K and DI water, followed by a thorough rinsing step using DI water. The DRIE process was performed using the SPTS Rapier OMEGA LPX system for a duration of 1.25 hrs to fabricate trenches. The dimensions of the trench patterns are as follows: 200 µm, 100 µm, 50 µm, 30 µm, 25 µm, 20 µm, 15 µm, 10 µm, and 5 µm, evenly spaced at intervals of 100 µm. $Ar/C_4F_8/SF_6/C_4F_8$ at flow rates of 200 sccm/200 sccm/1 sccm/120 sccm was used for the plasma strike step, lasting for 1 s at a pressure of 40 mTorr. Source 1 operated at a power of 2500 W, while source 2 operated at 1000 W. The Bosch process comprises a looped sequence involving polymer deposition (Dep), polymer removal (E1), and Si etching (E2). Dep occurs for 1.4 s at a pressure of 44 mTorr without bias, using $C_4F_8/SF_6/C_4F_8$ at flow rates of 360 sccm/1 sccm/95 sccm, with source 1 at 2500 W and source 2 at 1000 W. E1, lasting for 1.5 s at 25 mTorr, operates with source 1 at 2500 W and source 2 at 1000 W, maintaining a bias loop power between 75-110 W. This step uses $SF_6/C_4F_8$ at flow rates of 300 sccm/1 sccm. E2, executed for 1.9 s at 40 mTorr, utilizes $C_4F_8/SF_6$ at flow rates of 1 sccm/400 sccm with source 1 at 2500 W and source 2 at 1000 W with no bias. This cycle is repeated 850 times between Dep and E2, and the substrate temperature is maintained at 19 °C. To remove the resist material, a combination of acetone, isopropyl alcohol, and deionized water was used. Any resist remnants were eliminated using a PVA TePla microwave $O_2$ asher. Subsequently, the samples were loaded into a thermal furnace and held at a temperature of 1100 °C for a duration of 1.5 hr. During this time, an estimated oxide growth of 800 nm occurred[3,4], as calculated using this tool (https://cleanroom.byu.edu/oxidetimecalc). Following this thermal treatment, a buffered oxide etch process was applied. This entire sequence of thermal treatment and oxide etch was repeated once more for the desired results. The trench structures underwent coating using the deposition procedures previously outlined for planar film. The transition temperature measurements of the PEALD TiN in trenches were conducted using a four-wire measurement and a different cryogenic setup. The temperature steadily decreased from approximately 3K to 100mK within a 30-minute period. A small radiation shield was used to cover the trenches, preventing 3K radiation from affecting the film.

Following the deposition within the trenches, laser ablation was utilized to selectively eliminate the TiN film from the edges and lips of these trenches[5]. This process was undertaken to establish distinct regions for the purpose of channeling the current through the TiN coated within a series of electrically isolated trenches. The ablation procedure was conducted in an open-air environment using a diode-pumped ytterbium-doped potassium gadolinium tungstate (Yb:KGW) femtosecond laser. The laser operated at a fundamental wavelength of 1028 nm, which was subsequently converted to 343 nm through a harmonic module. The ablation process was carried out using a laser scanning system equipped with a 100 mm focal length telecentric f-Theta lens. A scanning velocity of 500 mm/s was applied, and the focused spot size measured approximately 15 µm, featuring a hatch line overlap of 10 µm and a laser pulse rate of 100 kHz. The combination of the 15 µm spot size and an average power of 1.3 watts resulted in a fluence of approximately 7.5 J/cm$^2$.

## S4. **Step Coverage**

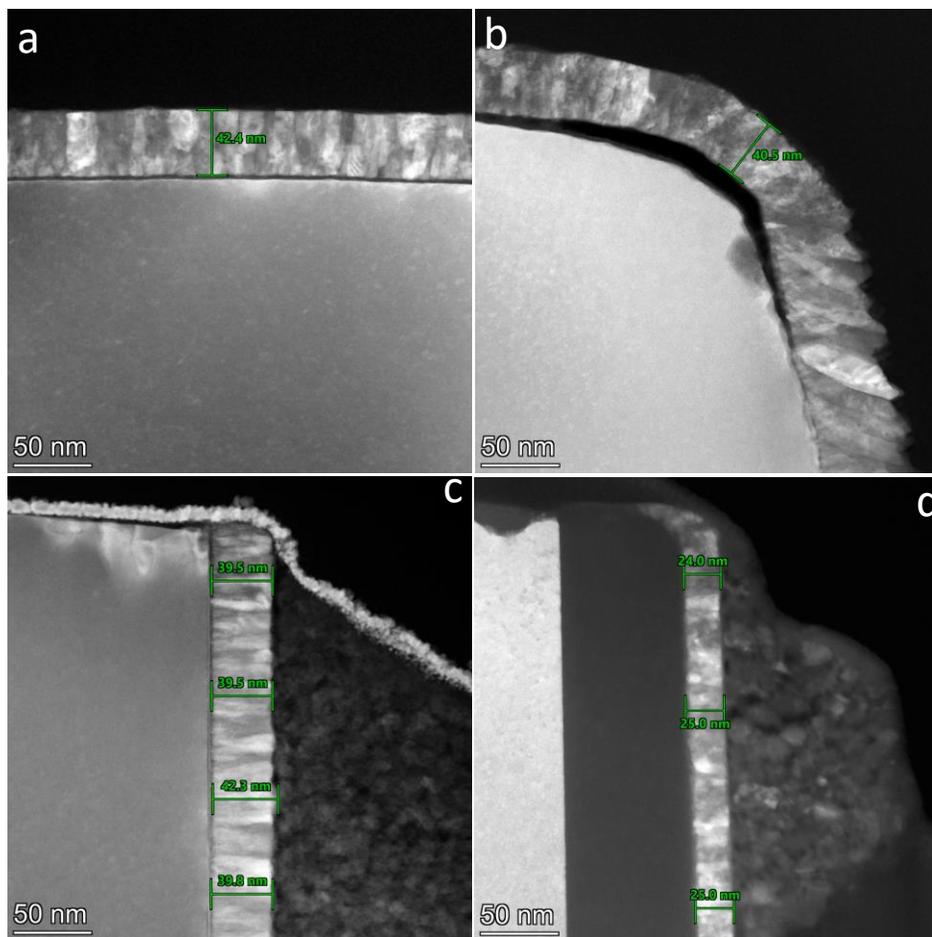

FIG. S4. illustrates the step coverage of TiN in Trench 9, displaying (a) the top, (b) the corner, (c) the sidewall, and (d) the bottom.
26

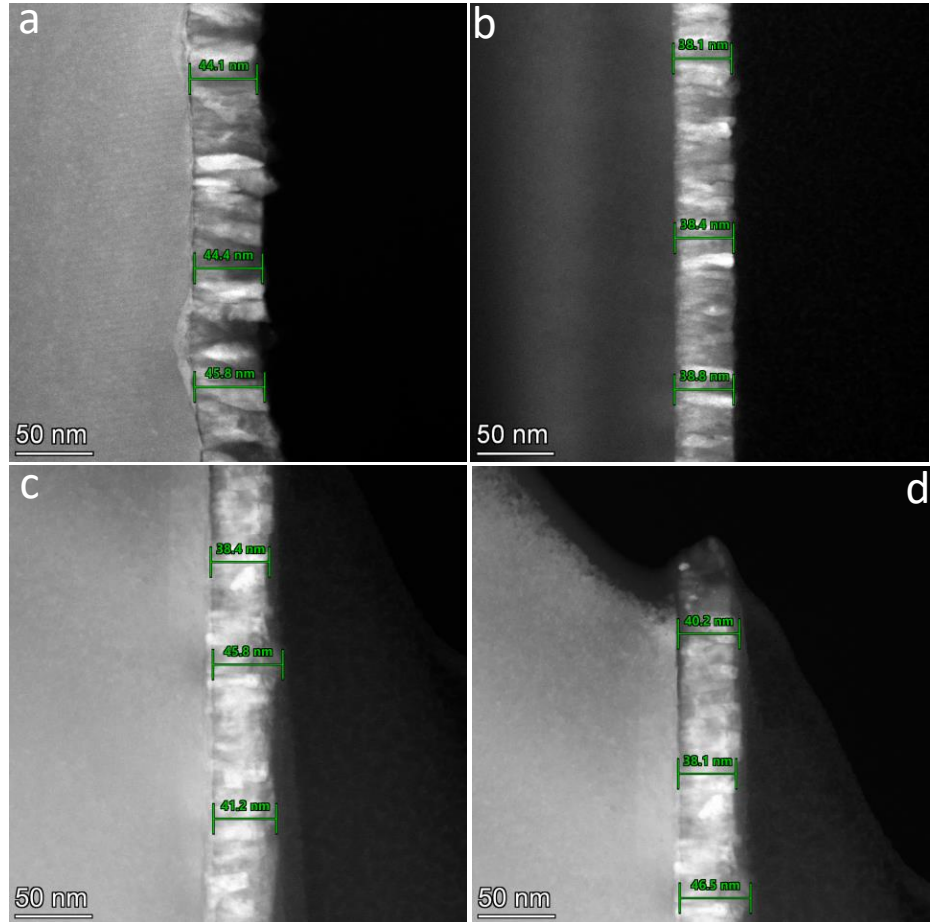

FIG. S5. illustrates the step coverage of TiN in Trench 7, it showcases (a) the upper location of the sidewall, (b) the lower location of the sidewall, (c) the bottom, and (d) an alternate location at the bottom.

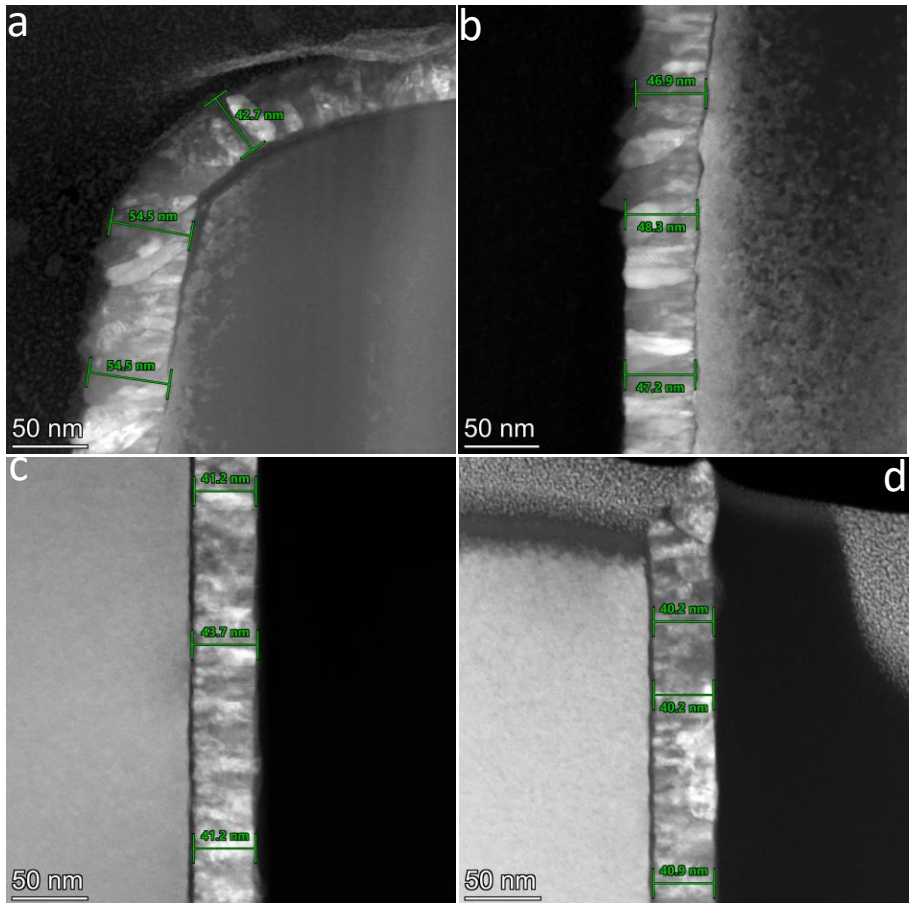

FIG. S6. illustrates the step coverage of TiN in Trench 1, displaying (a) the corner, (b) the upper location of the sidewall, (c) the bottom, and (d) an alternate location at the bottom.

## S5. Why does PEALD "Break" in 3D Features?

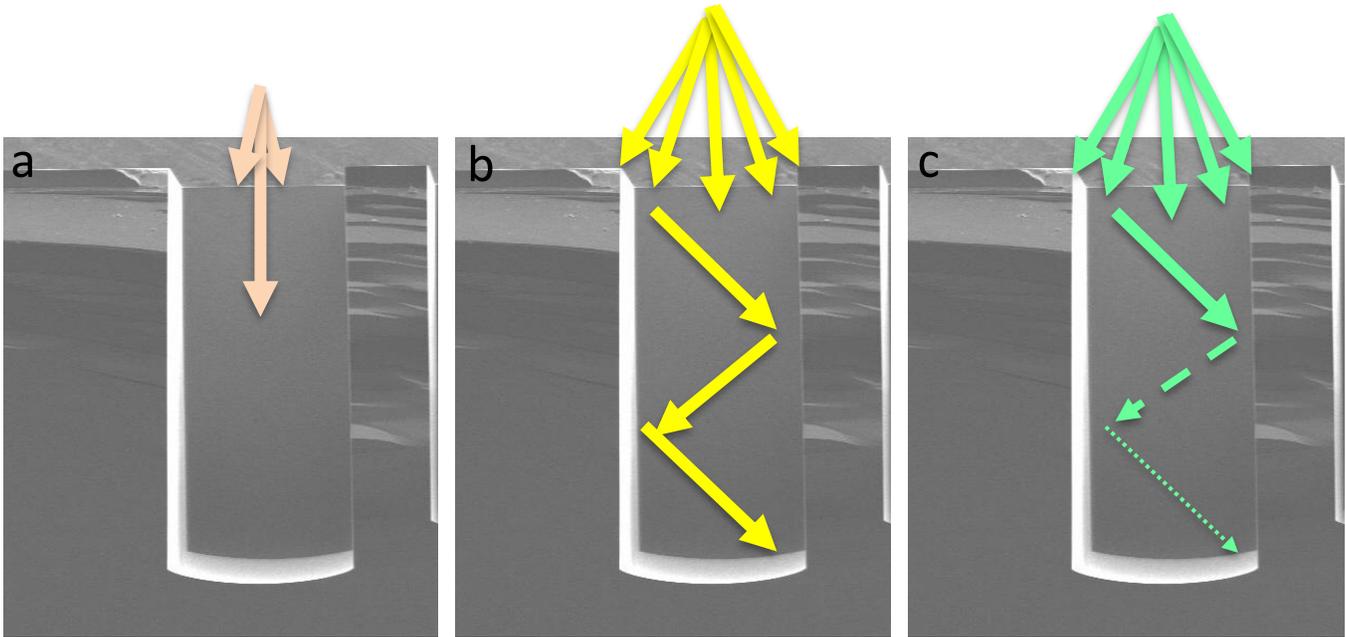

FIG. S6. The transport of ions and radicals to the trench bottom deteriorates as the AR of the trenches increases as described in (a) the ion flux to the sidewall is significantly lower compared to the field and occurs at a glancing angle, (b) the precursor flux endures through multiple bounces, and (c) reactive radicals (N*) exhibit a cosine distribution in their behavior.